\documentclass[final,twoside,twocolumn]{IEEEtran}
\usepackage{graphicx}
\usepackage{amsmath, amsfonts, amssymb, mathrsfs}
\usepackage{signal}
\usepackage{epic}
\usepackage{epsfig}
\usepackage[T1]{fontenc}

\newcommand{\supp}{\operatorname{supp} }

\newcommand{\bbZ}{{\mathbb{Z}}}
\newcommand{\bbR}{{\mathbb{R}}}
\newcommand{\bbC}{{\mathbb{C}}}

\newcommand{\expp}[1]{e^{#1}}

\newcommand{\ejt}{{{(e^{j\omega T})}}}

\newcommand{\bpsi}{\mbox{\boldmath{$\Psi$}}}
\newcommand{\bphi}{\mbox{\boldmath{$\Phi$}}}
\newcommand{\bgamma}{\mbox{\boldmath{$\Gamma$}}}

\newcommand{\bphil}{\mbox{\boldmath{$\phi$}}}
\newcommand{\bpgamma}{\mbox{\boldmath{$\gamma$}}}

\newcommand{\cls}{\mbox{span}}

\newcommand{\diag}{\mbox{diag}}

\newcommand{\ZZ}{{\mathbb Z}}

\newcommand{\bbb}{{{\bf B}}}

\newcommand{\bbf}{{{\bf F}}}
\newcommand{\bbg}{{{\bf G}}}

\newcommand{\bbi}{{{\bf I}}}
\newcommand{\bbm}{{{\bf M}}}
\newcommand{\bbq}{{{\bf Q}}}

\newcommand{\bbw}{{{\bf W}}}

\newcommand{\ba}{{{\bf a}}}

\newcommand{\bg}{{{\bf g}}}

\newcommand{\bx}{{{\bf x}}}
\newcommand{\by}{{{\bf y}}}

\newtheorem{example}{Example}
\newtheorem{proposition}{Proposition}

\newtheorem{problem}{Problem}

\newcommand{\beq}{\begin{equation}}
\newcommand{\eeq}{\end{equation}}

\begin{document}

\title{Recovering Signals from Lowpass Data
}

\author{Yonina C. Eldar,~\IEEEmembership{Senior~Member,~IEEE,} and Volker Pohl
\thanks{The authors are with the Department of Electrical Engineering,
        Technion -- Israel Institute of Technology,
        Haifa 32000, Israel,
                Phone: +972 4 829 3256, Fax:   +972 4 829 5757,
                e-mail: \{yonina,pohl\}@ee.technion.ac.il.}
\thanks{This work was supported in part by the Israel Science
Foundation under Grant no. 1081/07 and by the European Commission
in the framework of the FP7 Network of Excellence in Wireless
COMmunications NEWCOM++ (contract no. 216715). V. Pohl
acknowledges the support by the German Research Foundation (DFG)
under Grant PO~1347/1--1.} }

\markboth{Recovering Signals from Lowpass Data}{Eldar, Pohl}

\maketitle

\begin{abstract}
The problem of recovering a signal from its low frequency
components occurs often in practical applications due to the
lowpass behavior of many physical systems. Here we study in detail
conditions under which a signal can be determined from its
low-frequency content. We focus on signals in shift-invariant
spaces generated by multiple generators. For these signals, we
derive necessary conditions on the cutoff frequency of the lowpass
filter as well as necessary and sufficient conditions on the
generators such that signal recovery is possible. When the lowpass
content is not sufficient to determine  the signal, we propose
appropriate pre-processing that can improve the reconstruction
ability. In particular, we show that modulating the signal with
one or more mixing functions prior to lowpass filtering, can
ensure the recovery of the signal in many cases, and reduces the
necessary bandwidth of the filter.
\end{abstract}


\section{Introduction}
\label{sec:Intro}

Lowpass filters are prevalent in biological, physical and
engineering systems. In many scenarios, we do not have access to
the entire frequency content of a signal we wish to process, but
only to its low frequencies. For example, it is well known that
parts of the visual system exhibit lowpass nature: the neurons of
the outer retina have strong response to low frequency stimuli,
due to the relatively slow response of the photoreceptors. Similar
behavior is observed also in the cons and rods
\cite{Hung_Ciuffreda_02}. Another example is the lowpass nature of
free space wave propagation \cite{G05}. This limits the resolution
of optical image reconstruction to half the wave length. Many
engineering systems introduce lowpass filtering as well. One
reason is to allow subsequent sampling and digital signal
processing at a low rate.

Clearly if we have no prior knowledge on the original signal, and
we are given a lowpassed version of it, then we cannot recover the
missing frequency content. However, if we have prior knowledge on
the signal structure then it may be possible to interpolate it
from the given data. As an example, consider a signal $x$ that
lies in a shift-invariant (SI) space generated by a generator
$\phi$, so that $x(t)=\sum a_n\phi(t-nT)$ for some $T$. Even if
$x$ is not bandlimited, it can be recovered from the output of a
lowpass filter with cutoff frequency $\pi/T$ as long as the
Fourier transform $\hat{\phi}(\omega)$ of the generator is not
zero for all $\omega \in [-\pi/T,\pi/T)$
\cite{EM09,AldroubiGroechinger_01}.

The goal of this paper is to study in more detail under what
conditions a signal $x$ can be recovered from its low-frequency
content. Our focus is on signals that lie in SI spaces, generated
by multiple generators \cite{DDR94,GHM94,CE05}. Following a
detailed problem formulation in Section~\ref{sec:Problem}, we
begin in Section~\ref{sec:RecovCond} by deriving a necessary
condition on the cutoff frequency of the low pass filter (LPF) and
sufficient conditions on the generators such that $x$ can be
recovered from its lowpassed version. As expected, there are
scenarios in which recovery is not possible. For example, if the
bandwidth of the LPF is too small, or if one of the generators is
zero over a certain frequency interval and all of its shifts with
period $2\pi/T$, then recovery cannot be obtained. For cases in
which the recovery conditions are satisfied, we provide a concrete
method to reconstruct $x$ from the its lowpass frequency content
in Section~\ref{sec:RecovAlg}.

The next question we address is whether in cases in which the
recovery conditions are not satisfied, we can improve our ability
to determine the signal by appropriate pre-processing. In
Section~\ref{sec:Filter} we show that pre-processing with linear
time-invariant (LTI) filters does not help, even if we allow for a
bank of LTI filters. As an alternative, in
Section~\ref{sec:Mixers} we consider pre-processing by modulation.
Specifically, the signal $x$ is modulated by multiplying it with a
periodic mixing function prior to lowpass filtering. We then
derive conditions on the mixing function to ensure perfect
recovery. As we show, a larger class of signals can be recovered
this way. Moreover, by applying a bank of mixing functions, the
necessary cutoff frequency in each channel can be reduced. In
Section~\ref{sec:sparse} we briefly discuss how the results we
developed can be applied to sampling sparse signals in SI spaces
at rates lower than Nyquist. These ideas rely on the recently
developed framework for analog compressed sensing
\cite{ME08,E08,E08a}. In our setting, they translate to reducing
the LPF bandwidth, or the number of modulators. Finally,
Section~\ref{sec:Conclusions} summarizes and points out some open
problems.

Modulation architectures have been used previously in different
contexts of sampling. In \cite{GJ96} modulation was used in order
to obtain high-rate sigma-delta converters. More recently,
modulation has been used in order to sample sparse high bandwidth
signals at low rates \cite{LKDRBM07,ME09}. Our specific choice of
periodic functions is rooted in \cite{ME09} in which a similar
bank of modulators was used in order to sample multiband signals
at sub-Nyquist rates. Here our focus is on signals in general SI
spaces and our goal is to develop a broad framework that enables
pre-processing such as to ensure perfect reconstruction. We treat
signals that lie in a predefined subspace, in contrast to the
union of subspaces assumption used in the context of sparse signal
models. Our results can be used in practical systems that involve
lowpass filtering to pre-process the signal so that all its
content can be recovered from the received low-frequency signal
(without requiring a sparse signal model).

\section{Problem Formulation}
\label{sec:Problem}

\subsection{Notations}

We use the following notation: As usual, $\bbC^{N}$, $L^{2}$, and
$\ell^{2}$ denote the $N$-dimensional Euclidean space, the space
of square integrable function on the real line, and the space of
square summable sequences, respectively. All these spaces are
Hilbert spaces with the usual inner products. Throughout the paper
we write $\hat{x}$ for the Fourier transform of a function $x \in
L^{2}$:
\begin{equation*}
    \hat{x}(\omega)
    = \int^{\infty}_{-\infty} x(t)\, e^{-j \omega t}\, d t\,,
    \quad \omega \in \bbR.
\end{equation*}
The \emph{Paley-Wiener space} of functions in $L^{2}$ that are bandlimited to $[-B,B]$ will be denoted by $PW(B)$:
\begin{equation*}
    PW(B) = \{ x\in L^{2} : \hat{x}(\omega) = 0\ \text{for all}\ \omega\notin [-B,B]\},
\end{equation*}
and $P_{B}$ is the orthogonal projection $L^{2} \to PW(B)$ onto
$PW(B)$. Clearly, $P_{B}$ is a bounded linear operator on $L^{2}$.
We will also need the Paley-Wiener space of functions whose
inverse Fourier transform is supported on a compact interval, i.e.
\begin{equation*}
    \widehat{PW}(B) = \{ \hat{x}\in L^{2} : x(t) = 0\ \text{for all}\ t\notin [-B,B]\}.
\end{equation*}
For any $a\in\bbR$, the \emph{shift} (or translation) \emph{operator} $S_{a} : L^{2} \to L^{2}$ is defined by $(S_{a}x)(t) = x(t-a)$.

If $\{\phi_{k}\}_{k\in\mathcal{I}}$ is a set of functions in $L^{2}$ with an arbitrary index set $\mathcal{I}$ then
$\cls\left\{ \phi_{k} : k\in\mathcal{I} \right\}$
denotes the closed linear subspace of $L^{2}$ spanned by $\{\phi_{k}\}_{k\in\mathcal{I}}$.

\subsection{Problem Formulation}

\begin{figure}[h]
\begin{center}
\begin{picture}(26,6)(0,0)

\put(4,3){\st{$x(t)$}{\ar{\filter(8,5){$$}{\ar{\et{$y(t)$}}}}}}

\put(-16,-3.5){

\put(.1,.7){
    \put(26,5.5){\line(1,0){6}}
    \put(27,5.5){\line(0,1){1.7}}               
    \put(29,5.5){\line(0,1){2.3}}               
    \put(31,5.5){\line(0,1){1.7}}       
    \put(27,7.2){\line(1,0){4}}
    \put(26.8,4.5){\makebox(0,0){\footnotesize $-\pi/T_c$}}
    \put(31.2,4.5){\makebox(0,0){\footnotesize $\pi/T_c$}}
}
}

\end{picture}
\end{center}
\caption{Lowpass filtering of $x(t)$.} \label{fig:lpf}
\end{figure}
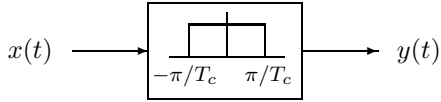

We consider the problem of recovering a signal $x(t)$, $t\in\bbR$
from its low-frequency content. Specifically, suppose that $x$ is
filtered by a LPF with cut off frequency $\pi/T_c$, as in
Fig.~\ref{fig:lpf}. We would like to answer the following
questions:
\begin{itemize}
\item What signals $x$ can be recovered from the output $y$
of the LPF? \item Can we perform preprocessing of $x$ prior to
filtering to ensure that $x$ can be recovered from $y$?
\end{itemize}
Filtering a signal $x \in L^{2}$ with a LPF with cutoff frequency
$\pi/T_{c}$ corresponds to a projection of $x$ onto the Paley
Wiener space $PW(\pi/T_{c})$.  Therefore we can write $y =
P_{\pi/T_{c}} x$.

Note, that we assume here that the output $y(t)$, $t\in\bbR$ is
analog. Since $y$ is a lowpass signal, an equivalent formulation
is to sample $y$ with period $T_{s}=1/f_{s}$ lower than the
Nyquist period $T_{c}$ to obtain the sequence of samples
$\{y[n]\}_{n \in \bbZ}$. The problem is then to recover $x(t)$,
$t\in\bbR$ from the samples $\{y[n]\}_{n \in \bbZ}$, as in
Fig.~\ref{fig:lpfs}.
 \setlength{\unitlength}{.09in}
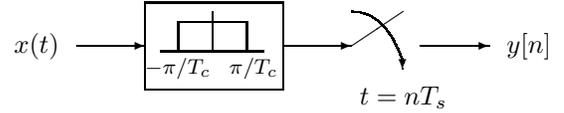
\begin{figure}[h]
\begin{center}
\begin{picture}(34,8)(0,0)

\put(0,0){ \put(4,4.5){\st{$x(t)$}{\ar{\filter(8,5){$$}{\ar}}}}
\put(-16.2,-2){

\put(.1,.7){
    \put(26,5.5){\line(1,0){6}}
    \put(27,5.5){\line(0,1){1.7}}               
    \put(29,5.5){\line(0,1){2.3}}               
    \put(31,5.5){\line(0,1){1.7}}       
    \put(27,7.2){\line(1,0){4}}
    \put(27.0,4.5){\makebox(0,0){\footnotesize $-\pi/T_c$}}
    \put(31.4,4.5){\makebox(0,0){\footnotesize $\pi/T_c$}}
}

}

\put(21,4.5){
   \put(0,0){\line(3,2){3}}
   \qbezier(0,2)(2,2)(3,-1)
   \put(2.95,-1.3){\vector(0,-1){0.05}}}
\put(24,1.5){\makebox(0,0){$t=nT_s$}}
\put(25,4.5){\ar{\et{$y[n]$}}}}

\end{picture}
\end{center}
\caption{Sampling of $x(t)$ after lowpass filtering.}
\label{fig:lpfs}
\end{figure}
Since $\{y[n]\}_{n\in\bbZ}$ uniquely determines $y$, the two
formulations are equivalent. For concreteness, we focus here on
the problem in which we are given $y(t)$, $t\in\bbR$ directly.
Thus, our emphasis is not on the sampling  rate, but rather on the
information content in the lowpass regime, regardless of the
sampling rate to follow.

Clearly, if $x$ is bandlimited to $[-\pi/T_{c},\pi/T_{c}]$, then
it can be recovered  from $y$. However, we will assume here that
$x$ is a general SI signal, not necessarily bandlimited. These
signals have the property that if $x(t)$ lies in a given SI space,
then so do all its shifts $(S_{kT} x)(t) = x(t - kT)$ by integer
multiples of some given $T$. Bandlimited signals are a special
class of SI signals. Indeed, if $x$ is bandlimited then so are all
its shifts $S_{kT}x$, $k\in\bbZ$ for a given $T$. In fact,
bandlimited signals have an even stronger property that all their
shifts $S_{a}x$ by any number $a\in\bbR$ are bandlimited.
Throughout, we assume that $x$ lies in a generally complex SI
space with multiple generators.

Let $\phi = \{\phi_{1},\dots,\phi_{N}\}$ be a given set of
functions in $L^{2}$ and let $T \in \bbR$ be a given real number.
Then the \emph{shift-invariant space} generated by $\phi$ is
formally defined as \cite{DDR94,GHM94,CE05}:
\begin{equation*}
    \mathcal{S}_{T}(\phi) = \cls\left\{ S_{kT}\phi_{n} : k\in\bbZ,1 \leq n \leq N \right\}.
\end{equation*}
The functions $\phi_{n}$ are referred to as the \emph{generators} of $\mathcal{S}_{T}(\phi)$.
Thus, every function $x \in \mathcal{S}_{T}(\phi)$ can be written as
\begin{equation}
\label{eq:si}
    x(t)
    = \sum_{n=1}^N \sum_{k \in \ZZ} a_{n}[k]\phi_{n}(t-kT),
    \quad t\in\bbR,
\end{equation}
where for each $1 \leq n \leq N$, $\{ a_{n}[k]\}_{k\in\ZZ}$ is an arbitrary sequence in $\ell^{2}$.
Examples of such SI spaces include multiband signals \cite{ME07} and spline functions \cite{S73b,EM09}. Expansions of the type (\ref{eq:si}) are also encountered in communication systems, when the analog signal is produced by pulse amplitude modulation.

In order to guarantee a unique and stable representation of any signal
in $\mathcal{S}_{T}(\phi)$ by sequences of coefficients $\{a_{n}[k]\}$, the generators
$\phi$ are typically chosen to form a \emph{Riesz basis} for
$\mathcal{S}_{T}(\phi)$. This means that there exist constants $\alpha>0$ and
$\beta<\infty$ such that
\begin{equation}
\label{eq:riesz}
    \alpha \|\ba\|^2
    \leq \left\| \sum_{n=1}^N
    \sum_{k \in \ZZ} a_{n}[k]\phi_{n}(t-kT)\right\|_{L^{2}}^2
    \leq \beta \|\ba\|^2,
\end{equation}
where $\|\ba\|^2=\sum_{n=1}^N \sum_{k \in \ZZ} |a_{n}[k]|^2$.
Condition (\ref{eq:riesz}) implies that any $x \in
\mathcal{S}_{T}(\phi)$ has a unique and stable representation in
terms of the sequences $\{a_{n}[k]\}_{k\in\bbZ}$. In particular,
it guarantees that the sequences $\{a_{n}[k]\}_{k\in\bbZ}$ can be
recovered from $x\in\mathcal{S}_{T}(\phi)$ by means of a linear
bounded operator.

By taking Fourier transforms in (\ref{eq:riesz}) it can be shown
that the generators $\phi$ form a Riesz basis\footnote{Here and in
the sequel, when we say that a set of generators $\phi$ form (or
generate) a basis, we mean that the basis functions are
$\{\phi_n(t-kT),k \in \ZZ,1 \leq n \leq N\}$.} if and only if
\cite{GHM94}
\begin{equation}
\label{eq:rc}
    \alpha \bbi \preceq \bbm_{\phi}(\omega) \preceq \beta \bbi,
    \quad \mbox{a.e. } \omega \in [-\pi/T,\pi/T].
\end{equation}
Here $\bbm_{\phi}(\omega)$ is called the \emph{Grammian} of the
generators $\phi = \{\phi_{1}, \dots, \phi_{N}\}$, and is the
$N\times N$ matrix
\begin{equation}
\label{eq:Mf}
    \bbm_{\phi}(\omega)
    = \left[\begin{array}{ccc}
            R_{\phi_1\phi_1}(\omega) & \ldots &
            R_{\phi_1\phi_N}(\omega)\\
            \vdots & \vdots &  \vdots \\
            R_{\phi_N\phi_1}(\omega) & \ldots & R_{\phi_N\phi_N}(\omega)
    \end{array} \right],
\end{equation}
where for any two generators $\phi_{i},\phi_{j}$ the function
$R_{\phi_{i}\phi_{j}}$ is given by
\begin{equation}
\label{eq:R}
    R_{\phi_{i}\phi_{j}}(\omega)
    = \sum_{k \in \ZZ}
    \overline{\hat{\phi}_{i}(\omega - 2k\,\tfrac{\pi}{T})}\, \hat{\phi}_{j}(\omega - 2k\, \tfrac{\pi}{T}).
\end{equation}
Note that the functions $R_{\phi_{i}\phi_{j}}$ are
$2\pi/T$-periodic. Therefore,
 condition \eqref{eq:rc} is equivalent to $\alpha \bbi \preceq \bbm_{\phi}(\omega - a) \preceq \beta \bbi$ for every arbitrary real number $a$. We will need in particular the case $a = \pi/T$, for which the entries of the matrix $\bbm_{\phi}(\omega-a)$ are
\begin{equation}
\label{eq:R2}
    R_{\phi_{i}\phi_{j}}(\omega-\tfrac{\pi}{T})
    = \sum_{k \in \ZZ}
    \overline{\hat{\phi}_{i}(\omega - [2k+1]\,\tfrac{\pi}{T})}\, \hat{\phi}_{j}(\omega - [2k+1]\, \tfrac{\pi}{T}).
\end{equation}

\section{Recovery Conditions}
\label{sec:RecovCond}

The first question we address is whether we can recover
$x\in\mathcal{S}_{T}(\phi)$ of the form (\ref{eq:si}) from the
output $y = P_{\pi/T_{c}}\, x$ of a LPF with cutoff frequency
$\pi/T_c$, assuming that the generators $\phi$ satisfy
(\ref{eq:rc}). We further assume that the generators are not
bandlimited to $\pi/T_c$, namely that they have energy outside the
frequency interval $[-\pi/T_{c},\pi/T_{c}]$. We will provide
conditions on the generators $\phi$ and on the bandwidth of the
LPF such that $x$ can be recovered from $y$. As we show, even if
the generators $\phi$ are not bandlimited, $x$ can often be
determined from $y$.

First we note that in order to recover $x\in\mathcal{S}_{T}(\phi)$ from the lowpass signal $y = P_{\pi/T_{c}}\, x$ it is sufficient to recover the sequences $\{a_{n}[k]\}_{k\in\bbZ}$, $1 \leq n \leq N$ because the generators $\phi$ are assumed to be known.
The output of the LPF can be written as
\begin{eqnarray*}
    y(t) = (P_{\pi/T_{c}}\, x)(t)
    = \sum_{n=1}^N \sum_{k \in \ZZ} a_{n}[k]\, \psi_{n}(t-kT)
\end{eqnarray*}
where $\psi_{n} := P_{\pi/T_{c}}\, \phi_{n}$ denotes the lowpass
filtered generator $\phi_{n}$, and the sum on the right-hand side
converges in $L^{2}$ since $P_{\pi/T_{c}}$ is bounded. Therefore,
we immediately have the following observation: The sequences
$\{a_{n}[k]\}_{k\in\bbZ}$, $1 \leq n \leq N$ can be recovered from
$y$ if $\psi$ forms a Riesz basis for $\mathcal{S}_{T}(\psi)$.
This is equivalent to the following statement.

\begin{proposition}
\label{lem:RecovCond} Let $\phi = \{\phi_{1},\dots,\phi_{N}\}$ be
a set generators, and let $\psi_{n} = P_{\pi/T_{c}}\phi_{n}$, $1
\leq n \leq N$ be the lowpass filtered generators where
$\pi/T_{c}$ is the bandwidth of the LPF. Then the signal $x \in
\mathcal{S}_{T}(\phi)$ can be recovered from the observations $y =
P_{\pi/T_{c}} x$ if the Grammian $\mathbf{M}_{\psi}(\omega)$
satisfies \eqref{eq:rc} for some $0 < \alpha \leq \beta < \infty$.
\end{proposition}

\begin{example}
\label{ex:sinc_N1}
We consider the case of one generator ($N=1$)
\begin{equation}
\label{equ:Sinc_Example}
    \phi_{1}(t) = \left\{\begin{array}{ll}
        1/(2 D), &\quad t\in [-D,D] \\
        0,     &\quad t\notin [-D,D]
    \end{array}\right.
\end{equation}
for some $D>0$. The Fourier transform of this generator is
$\hat{\phi}_{1}(\omega) = \sin(\omega D)/(\omega D)$ which becomes
zero at $\omega = k \pi/D$ for all $k = \pm1, \pm 2, \dots$. We
assume that $D/T$ is not an integer. Then one can easily see that
this generator satisfies \eqref{eq:rc}, i.e. there exists $\alpha,
\beta$ such that
\begin{eqnarray}
\label{equ:Ex1_1}
    0 < \alpha \leq
    \sum_{k \in \bbZ}
    \left| \frac{\sin(\omega D - 2\pi k D/T)}
                {\omega D - 2\pi k D/T }
    \right|^{2}
    \leq \beta < \infty
\end{eqnarray}
for all $\omega\in [-\pi/T,\pi/T]$. The lower bound follows from
the assumption that $D/T$ is not an integer, so that all the
functions in the above sum have no common zero in
$[-\pi/T,\pi/T]$. The upper bound $\beta$ follows from
\begin{multline*}
    \sum_{k\in \bbZ} \left| \frac{\sin(\omega D - 2\pi k D/T)}{\omega D - 2\pi k D/T} \right|^{2}
    \leq \sum_{k \in \bbZ} \frac{1}{\left| \omega D - 2\pi k D/T \right|^{2} }\\
    \leq \left( \frac{T}{\pi D} \right)^{2} \left[ 1 + 2\sum^{\infty}_{k=1} \frac{1}{(2 k - 1)^{2}} \right]
    \leq \left( \frac{2\, T}{\pi D} \right)^{2}
\end{multline*}
using that $|\omega\, D - 2\pi k D/T| \geq \pi D/T(2 |k| -1)$ for all $k = \pm 1, \pm2, \dots$ and all $\omega \in [-\pi/T,\pi/T]$.

Assume now that the LPF has cutoff frequency $\pi/T_{c} = \pi/T$.
Then the Fourier transform $\hat{\psi}_{1}$ of the filtered
generator $\psi_{1} = P_{\pi/T}\, \phi_{1}$ will satisfy a
relation like \eqref{equ:Ex1_1} only if $D\leq T$, i.e. only if
$\hat{\phi}_{1}$ has no zero in $[-\pi/T,\pi/T]$. In cases where
$D > T$ the cutoff frequency has to be larger in order to allow a
recovery of the original signal. One easily sees that the cutoff
frequency of the LPF has to lie at least $\pi/T - \pi/D$ above
$\pi/T$ in order that $\hat{\psi}_{1}$ will satisfy a relation
similar to \eqref{equ:Ex1_1}. In this case, the shifts
$\hat{\psi}_{1}(\omega \pm 2\pi/T)$ compensate for the zero of
$\hat{\psi}_{1}(\omega)$ in the sum \eqref{equ:Ex1_1}. Thus for
cutoff frequencies $\pi/T_{c} \geq 2\pi/T - \pi/D$ a recovery of
the signal $x$ from the LPF signal $y$ will be possible.
\end{example}

The previous example illustrates that the question whether $\psi$
forms a Riesz basis for $\mathcal{S}_{T}(\psi)$ depends on the
given generators $\phi$ and on the bandwidth $\pi/T_{c}$ of the
LPF. The next proposition derives a necessary condition on the
required bandwidth $\pi/T_{c}$ of the LPF such that $\psi$ can be
a Riesz basis for $\mathcal{S}_{T}(\psi)$.

\begin{proposition}
\label{lem:MinBW}
Let $\phi = \{\phi_{1},\dots,\phi_{N}\}$ be a Riesz basis for the space $\mathcal{S}_{T}(\phi)$ and let $\psi_{n} = P_{\pi/T_{c}} \phi_{n}$ with $1 \leq n \leq N$. Then a necessary condition for $\psi = \{\psi_{1},\dots,\psi_{N}\}$ to be a Riesz basis for $\mathcal{S}_{T}(\psi)$ is that $\pi/T_{c} \geq N \pi/T$.
\end{proposition}

\begin{proof}
We consider the Grammian $\bbm_{\psi}(\omega)$ whose entries are
equal to
\begin{equation*}
    R_{\psi_{i}\psi_{j}}(\omega)
    = \sum_{|k| \leq \frac{1}{2}(\frac{T}{T_{c}} + 1)}
    \overline{\hat{\psi}_{i}(\omega - k\,\tfrac{2\pi}{T})}\, \hat{\psi}_{j}(\omega - k\, \tfrac{2\pi}{T}).
\end{equation*}
All other terms in the generally infinite sum (cf. \eqref{eq:R})
are identically zero since $\hat{\psi}_{n}(\omega)$ is bandlimited
to $[-\pi/T_{c},\pi/T_{c}]$. This Grammian can be written as
$\bbm_{\psi}(\omega) = \bpsi^{*}(\omega)\bpsi(\omega)$ with
\begin{multline}
\label{eq:psim}
    \bpsi(\omega) =\\
    \left[ \begin{array}{ccc}
    \hat{\psi}_1(\omega + [L_{0}+1] \tfrac{2\pi}{T})  & \ldots & \hat{\psi}_N(\omega + [L_{0}+1] \tfrac{2\pi}{T})\\
    \hat{\psi}_1(\omega + L_{0} \tfrac{2\pi}{T})  & \ldots & \hat{\psi}_N(\omega + L_{0} \tfrac{2\pi}{T})\\
    \vdots &&\vdots \\
    \hat{\psi}_1(\omega) & \ldots & \hat{\psi}_N(\omega)  \\
    \vdots &&\vdots \\
    \hat{\psi}_1(\omega - L_{0} \tfrac{2\pi}{T} ) & \ldots & \hat{\psi}_N(\omega - L_{0} \tfrac{2\pi}{T} )\\
    \hat{\psi}_1(\omega - [L_{0}+1] \tfrac{2\pi}{T} ) & \ldots & \hat{\psi}_N(\omega - [L_{0}+1] \tfrac{2\pi}{T} )
    \end{array} \right]
\end{multline}
where $L_{0}$ is the largest integer such that $L_{0} \leq
(T/T_{c} - 1)/2$. Since every $\hat{\psi}_{n}(\omega)$ is banded
to $[-\pi/T_{c},\pi/T_{c}]$, the first and the last row of this
matrix are identically zero for some $\omega \in [-\pi/T,\pi/T]$.
At these $\omega$'s, the matrix $\bpsi(\omega)$ has effectively $L
= 2 L_{0} + 1$ rows and $N$ columns, and it holds that $L \leq
T/T_{c}$. Since $\bbm_{\psi}(\omega) =
\bpsi^{*}(\omega)\bpsi(\omega)$, the Grammian can have full rank
for every $\omega\in [-\pi/T,\pi/T]$ only if $L \geq N$, i.e. only
if $\pi/T_{c} \geq N \pi/T$.
\end{proof}

The necessary condition on the bandwidth of the LPF given in the
previous proposition is not sufficient, in general. However, given
a bandwidth $\pi/T_{c}$ which satisfies the necessary condition of
Proposition~\ref{lem:MinBW}, sufficient conditions on the
generators $\phi$ can be derived such that the lowpass filtered
generators $\psi$ form a Riesz basis for $\mathcal{S}_{T}(\psi)$,
i.e. such that $x$ can be recovered from $y$.

\begin{proposition}
\label{lem:SuffCond} Let $\phi = \{\phi_{1},\dots,\phi_{N}\}$ be a
Riesz basis for $\mathcal{S}_{T}(\phi)$ and let $\psi_{n} =
P_{\pi/T_{c}}\phi_{n}$ for $1 \leq n \leq N$ with $\pi/T_{c} \geq
N\, \pi/T$. Denote by $L$ the largest integer such that $L\leq
T/T_{c}$. If $L = 2 L_{0} + 1$ is an odd number, then we define
the $L\times N$ matrix $\bphi_{L}(\omega)$ by
\begin{multline}
\label{eq:phim}
    \bphi_{L}(\omega) =
    \left[ \begin{array}{ccc}
    \hat{\phi}_1(\omega + 2\, L_{0} \tfrac{\pi}{T})  & \ldots & \hat{\phi}_N(\omega + 2\, L_{0} \tfrac{\pi}{T})\\
    \vdots &&\vdots \\
    \hat{\phi}_1(\omega + 2\, \tfrac{\pi}{T})  & \ldots & \hat{\phi}_N(\omega + 2\, \tfrac{\pi}{T})\\
    \hat{\phi}_1(\omega) & \ldots & \hat{\phi}_N(\omega)  \\
    \hat{\phi}_1(\omega - 2\, \tfrac{\pi}{T})  & \ldots & \hat{\phi}_N(\omega - 2\, \tfrac{\pi}{T})\\
    \vdots &&\vdots \\
    \hat{\phi}_1(\omega - 2\, L_{0} \tfrac{\pi}{T} ) & \ldots & \hat{\phi}_N(\omega - 2\, L_{0} \tfrac{\pi}{T} )
    \end{array} \right].
\end{multline}
For $L = 2 L_{0}$ even, we define
\begin{multline}
\label{eq:phim2}
    \bphi_{L}(\omega) = \\
    \left[ \begin{array}{ccc}
    \hat{\phi}_1(\omega + [2 L_{0}- 1] \tfrac{\pi}{T})  & \ldots & \hat{\phi}_N(\omega + [2 L_{0}- 1] \tfrac{\pi}{T})\\
    \vdots &&\vdots \\
    \hat{\phi}_1(\omega + \tfrac{\pi}{T}) & \ldots & \hat{\phi}_N(\omega + \tfrac{\pi}{T})\\
    \hat{\phi}_1(\omega - \tfrac{\pi}{T}) & \ldots & \hat{\phi}_N(\omega - \tfrac{\pi}{T})\\
    \vdots &&\vdots \\
    \hat{\phi}_1(\omega - [2 L_{0}- 1] \tfrac{\pi}{T} ) & \ldots & \hat{\phi}_N(\omega - [2 L_{0}- 1] \tfrac{\pi}{T} )
    \end{array} \right].
\end{multline}

If there exists a constant $\alpha > 0$ such that
\begin{equation}
\label{equ:LemSuffCond}
    \bbm_{L}(\omega) := \bphi^{*}_{L}(\omega)\,\bphi_{L}(\omega) \succeq \alpha \bbi
    \quad\text{a.e.}\ \omega \in [-\tfrac{\pi}{T},\tfrac{\pi}{T}]
\end{equation}
then $\psi = \{\psi_{1},\dots,\psi_{N}\}$ forms a Riesz basis for $\mathcal{S}_{T}(\psi)$.

Moreover, if $T/T_{c}$ is an integer, then condition \eqref{equ:LemSuffCond} is also necessary for $\psi$ to be a Riesz basis for $\mathcal{S}_{T}(\psi)$.
\end{proposition}

When $\pi/T_{c} \to \infty$, i.e.  $L \to \infty$, the matrix
$\bbm_{L}(\omega)$ reduces to $\bbm_{\phi}(\omega)$ of
(\ref{eq:Mf}), which by definition satisfies \eqref{eq:rc}.
However, since for the calculation of the entries of
$\bbm_{L}(\omega)$ we are only summing over a partial set of the
integers, we are no longer guaranteed that $\bbm_{L}(\omega)$
satisfies the lower bound of \eqref{eq:rc}.

The requirements of Proposition~\ref{lem:SuffCond} imply that $L
\geq N$. Consequently, the matrix $\bbm_{L}(\omega) =
\bphi^{*}_{L}(\omega)\,\bphi_{L}(\omega)$ is positive definite for
almost all $\omega \in [-\pi/T,\pi/T]$ if and only if
$\bphi_{L}(\omega)$ has full column rank for almost all $\omega
\in [-\pi/T,\pi/T]$.

Note that Example~\ref{ex:sinc_N1} shows that
\eqref{equ:LemSuffCond} is not necessary, in general: With $T < D
< 2T$ and a cutoff frequency of $\pi/T_{c} > 2\pi/T - \pi/D$, the
corresponding $\psi$ form a Riesz basis for
$\mathcal{S}_{T}(\psi)$. However, it can easily be verified that
\eqref{equ:LemSuffCond} is not satisfied.

\begin{proof}
We consider the case of $L$ being odd. It has to be shown that the
Grammian $\bbm_{\psi}(\omega)$ satisfies  \eqref{eq:rc}. Since
$NT_c \leq T$,  the Grammian can be written as
$\bbm_{\psi}(\omega) = \bpsi^{*}(\omega)\bpsi(\omega)$ with
$\bpsi(\omega)$  defined by \eqref{eq:psim}. Next $\bpsi(\omega)$
is written as $\bpsi(\omega) = \bpsi_{L}(\omega) +
\bpsi_{\perp}(\omega)$ where $\bpsi_{\perp}(\omega)$ is the $(2
L_{0} +1) \times N$ matrix whose first and last row coincide with
those of $\bpsi(\omega)$ and whose other rows are identically
zero. Similarly $\bpsi_{L}(\omega)$ denotes the matrix whose first
and last row is identically zero and whose remaining rows coincide
with those of $\bpsi(\omega)$. Since $\psi_{n}(\omega) =
\phi_{n}(\omega)$ for all $\omega \in [-\pi/T_{c},\pi/T_{c}]$ and
for every $1 \leq n \leq N$, we have that
$\bpsi^{*}_{L}(\omega)\,\bpsi_{L}(\omega) =
\bphi^{*}_{L}(\omega)\,\bphi_{L}(\omega)$. Therefore,
\begin{multline}
\label{equ:Proof_SuffCond_01}
    \bbm_{\psi}(\omega)
    = \bpsi^{*}_{L}(\omega)\,\bpsi_{L}(\omega) + \bpsi^{*}_{\perp}(\omega)\,\bpsi_{\perp}(\omega)\\
    + \bpsi^{*}_{L}(\omega)\,\bpsi_{\perp}(\omega) + \bpsi^{*}_{\perp}(\omega)\,\bpsi_{L}(\omega)\\
    = \bphi^{*}_{L}(\omega)\,\bphi_{L}(\omega) + \bpsi^{*}_{\perp}(\omega)\,\bpsi_{\perp}(\omega)
\end{multline}
since by the definition of $\bpsi_{L}(\omega)$ and
$\bpsi_{\perp}(\omega)$, we obviously have that
$\bpsi^{*}_{L}(\omega)\,\bpsi_{\perp}(\omega) \equiv 0$ and
$\bpsi^{*}_{\perp}(\omega)\,\bpsi_{L}(\omega) \equiv 0$. Now it
follows from \eqref{equ:Proof_SuffCond_01} that for every $\bx \in
\bbC^{N}$
\begin{eqnarray*}
    \bx^{*}\, \bbm_{\psi}(\omega)\, \bx
    &=&    \|\bphi_{L}(\omega)\,\bx\|^{2}_{\bbC^{N}} + \|\bpsi_{\perp}(\omega)\,\bx\|^{2}_{\bbC^{N}}\\
    &\geq& \|\bphi_{L}(\omega)\,\bx\|^{2}_{\bbC^{N}}
    =      \bx^{*}\, \bphi^{*}_{L}(\omega)\bphi_{L}(\omega)\, \bx
    \geq \alpha,
\end{eqnarray*}
where the last inequality follows from \eqref{equ:LemSuffCond}.
This shows that the Grammian $\bbm_{\psi}(\omega)$ is lower
bounded as in \eqref{eq:rc}. The existence of an upper bound for
$\bbm_{\psi}(\omega)$ is trivial since $\bbm_{\psi}(\omega)$ has
finite dimensions.

Assume now that $T/T_{c}$ is an (odd) integer. In this case $L_{0}
= (T/T_{c} - 1)/2$ and it can easily be verified that the matrix
$\bpsi_{\perp}(\omega)$ is identically zero. From
\eqref{equ:Proof_SuffCond_01}, $\bbm_{\psi}(\omega) =
\bphi^{*}_{L}(\omega)\,\bphi_{L}(\omega) = \bbm_{L}(\omega)$ which
shows that if the Grammian $\bbm_{\psi}(\omega)$ satisfies
\eqref{eq:rc} then $\bphi_{L}(\omega)$ satisfies
\eqref{equ:LemSuffCond}. This proves that \eqref{equ:LemSuffCond}
is also necessary for $\psi$ to be a Riesz basis for
$\mathcal{S}_{T}(\psi)$.

The case of $L$ even follows from the same arguments but starting
with expression \eqref{eq:R2} for the entries of the Grammian
instead of \eqref{eq:R}. Therefore, the details are omitted.
\end{proof}

\begin{example}
\label{ex:sinc_N2} We consider an example with two generators
(N=2) which both have the form as in Example~\ref{ex:sinc_N1},
with different values for $D$, i.e.
\begin{equation*}
    \phi_{i}(t) = \left\{\begin{array}{ll}
        1/(2 D_{i}), &\quad t\in [-D_{i},D_{i}] \\
        0,     &\quad t\notin [-D_{i},D_{i}]
    \end{array}\right.
    \quad i=1,2
\end{equation*}
with Fourier transforms $\hat{\phi}_{i}(\omega) = \sin(\omega
D_{i})/(\omega D_{i})$. As in Example~\ref{ex:sinc_N1} we assume
that $D_{i}/T$ are not integers and that $D_{1} \neq D_{2}$. Under
these conditions, the Grammian $\bbm_{\phi}(\omega)$ of $\phi =
\{\phi_{1},\phi_{2}\}$ satisfies \eqref{eq:rc}. To see this, we
consider the determinant of $\bbm_{\phi}(\omega)$ for some
arbitrary but fixed $\omega\in [-\pi/T,\pi/T]$:
\begin{multline}
\label{eq:ExSincN2_Det1}
    \det[\bbm_{\phi}(\omega)]
    = \sum_{k\in\bbZ} \left| \hat{\phi}_{1}(\omega - k \tfrac{2\pi}{T}) \right|^{2}
      \sum_{k\in\bbZ} \left| \hat{\phi}_{2}(\omega - k \tfrac{2\pi}{T}) \right|^{2}\\
    - \left( \sum_{k\in\bbZ} \hat{\phi}_{1}(\omega - k \tfrac{2\pi}{T})\, \hat{\phi}_{2}(\omega - k \tfrac{2\pi}{T}) \right)^{2}.
\end{multline}
We know from Example~\ref{ex:sinc_N1}, that the first term on the
right hand side is lower bounded by some constant
$\alpha_{1}\alpha_{2} > 0$. Moreover, the Cauchy-Schwarz
inequality shows that the second term on the right-hand side is
always smaller or equal than the first term with equality only if
the two sequences
\begin{eqnarray*}
    \{ \hat{\phi}_{i}(\omega - k \tfrac{2\pi}{T}) \}_{k\in\bbZ},
    \quad i = 1,2
\end{eqnarray*}
are linearly dependent. However, since $D_{1} \neq D_{2}$, it is
not hard to verify that these two sequences are linearly
independent. Consequently $\det[\bbm_{\phi}(\omega)] > 0$ for all
$\omega\in [-\pi/T,\pi/T]$ which shows that $\bbm_{\phi}(\omega)$
satisfies the lower bound of \eqref{eq:rc}. That
$\bbm_{\phi}(\omega)$ satisfies also the upper bound in
\eqref{eq:rc} follows from a similar calculation as in
Example~\ref{ex:sinc_N1} using that $|\hat{\phi}_{i}(\omega)|$
deceases proportional to $1/\omega$ as $|\omega|\to\infty$.

Assume now that the bandwidth of the LPF satisfies $2 \pi/T \leq
\pi/T_{c} < 3 \pi/T$. In this case the matrix $\bphi_{L}(\omega)$
of Proposition~\ref{lem:SuffCond} is given by
\begin{equation*}
    \bphi_{L}(\omega) = \\
    \left[ \begin{array}{cc}
    \hat{\phi}_1(\omega + \tfrac{\pi}{T}) & \hat{\phi}_2(\omega + \tfrac{\pi}{T})\\
    \hat{\phi}_1(\omega - \tfrac{\pi}{T}) & \hat{\phi}_2(\omega - \tfrac{\pi}{T})\\
    \end{array} \right],
\end{equation*}
and the determinant of $\bbm_{L}(\omega) :=
\bphi^{*}_{L}(\omega)\,\bphi_{L}(\omega)$ becomes
\begin{multline}
\label{eq:ExSincN2_Det2}
    \det[\bbm_{L}(\omega)]
    = \sum_{k = \pm 1} \left| \hat{\phi}_{1}(\omega - k \tfrac{\pi}{T}) \right|^{2}
      \sum_{k = \pm 1} \left| \hat{\phi}_{2}(\omega - k \tfrac{\pi}{T}) \right|^{2}\\
    - \left( \sum_{k = \pm 1} \hat{\phi}_{1}(\omega - k \tfrac{\pi}{T})\, \hat{\phi}_{2}(\omega - k \tfrac{\pi}{T}) \right)^{2}.
\end{multline}
This expression is similar to \eqref{eq:ExSincN2_Det1} and the
same arguments show that $\det[\bbm_{L}(\omega)] > 0$ for all
$\omega \in [-\pi/T,\pi/T]$. Namely, since $D_{i}/T$ are not
integers, the functions $\hat{\phi_{i}}(\omega + \pi/T)$ and
$\hat{\phi_{i}}(\omega - \pi/T)$ have no common zero such that the
first term on the right hand side of \eqref{eq:ExSincN2_Det2} is
lower bounded by some $\alpha_{1}\alpha_{2} > 0$. The
Cauchy-Schwarz inequality implies that the second term is always
smaller than the first one.

We conclude that $\bphi_{L}(\omega)$ satisfies the condition of
Proposition~\ref{lem:SuffCond}, so that the signal $x$ can be
recovered from its low frequency components $y = P_{\pi/T_{c}}x$.
\end{example}

If for a certain bandwidth $\pi/T_{c}$ of the LPF the generators
$\phi$ satisfy the conditions of Proposition~\ref{lem:SuffCond}
then the signal $x$ can be recovered from $y = P_{\pi/T_{c}} x$.
However, if the generators $\phi$ do not satisfy these conditions,
then there exists in principle two ways to enable recovery of $x$:
\begin{itemize}
\item Increasing the bandwidth of the LPF.
 \item Pre-process $x$
before lowpass filtering, i.e. modify the generators $\phi$.
\end{itemize}
It is clear that for a given set $\phi =
\{\phi_{1},\dots\phi_{N}\}$ of generators an increase of the LPF
can only increase the "likelihood" that the matrix
$\bphi_{L}(\omega)$ of Proposition~\ref{lem:SuffCond} will have
full column rank. This is because enlarging $\pi/T_{c}$ increases
the number $L$ i.e. it adds additional rows to the matrix which
can only enlarge the column rank of $\bphi_{L}(\omega)$.
Pre-processing of $x$ will be discussed in detail in
Sections~\ref{sec:Filter} and \ref{sec:Mixers}.

\section{Recovery Algorithm}
\label{sec:RecovAlg}

We now describe a simple method to reconstruct the desired signal
$x$ from its low frequency components. This method is used in
later sections to show how pre-processing of the signal $x$ may
facilitate its recovery. Throughout this section, we assume that
the bandwidth $\pi/T_{c}$ of the LPF satisfies the necessary
condition of Proposition~\ref{lem:MinBW}, and that the generators
satisfy the sufficient condition of
Proposition~\ref{lem:SuffCond}.

Taking the Fourier transform of (\ref{eq:si}), we see that every $x \in \mathcal{S}_{T}(\phi)$ can be expressed in the Fourier domain as
\begin{equation}
\label{eq:xeq}
    \hat{x}(\omega)
    = \sum_{n=1}^N \hat{a}_{n}(e^{j\omega T})\,\hat{\phi}_{n}(\omega),
    \quad \omega\in\bbR
\end{equation}
where
\begin{equation*}
    \hat{a}_{n}(e^{j\omega T}) = \sum_{k \in \ZZ} a_{n}[k] e^{-j\omega 
k T}
\end{equation*}
is the $2\pi/T$-periodic discrete time Fourier transform of the
sequence $\{a_{n}[k]\}_{k\in\bbZ}$ at frequency $\omega T$.
Denoting by $\hat{\ba}\ejt$ the vector whose $n$th element is
equal to $\hat{a}_{n}\ejt$ and by $\hat{\bphil}(\omega)$ the
vector whose $n$th element is equal to $\hat{\phi}_{n}(\omega)$ we
can write (\ref{eq:xeq}) in vector form as
\[\hat{x}(\omega) = \hat{\bphil}^{T}(\omega)\,\hat{\ba}\ejt.\]

 The
Fourier transform of the LPF output $y = P_{\pi/T_{c}}\, x$ is
bandlimited to $\pi/T_{c}$, and for all $\omega \in
[-\pi/T_{c},\pi/T_{c}]$ we have $\hat{y}(\omega) =
\hat{x}(\omega)$. Therefore
\begin{equation}
\label{eq:seq}
    \hat{y}(\omega) = \hat{\bphil}^T(\omega)\,\hat{\ba}\ejt\,,
    \quad  \omega \in [-\tfrac{\pi}{T_{c}},\tfrac{\pi}{T_{c}}].
\end{equation}
For every $\omega \in [-\pi/T_{c},\pi/T_{c}]$,  \eqref{eq:seq}
describes an equation for the $N$ unknowns $\hat{a}_{n}\ejt$.
Clearly, one equation is not sufficient to recover the length-$N$
vector $\hat{\ba}\ejt$; we need at least $N$ equations.
 However, since
according to Proposition~\ref{lem:MinBW} the bandwidth of the LPF
has to be at least $\pi/T_{c} \geq N \pi/T$, we can form more
equations from the given data by noting that $\hat{\ba}$ is
periodic with period $2\pi/T$, while $\hat{\bphil}$, and
consequently $\hat{y}$, are generally not. Specifically, let
$\omega_{0} \in [-\pi/T,\pi/T]$ be an arbitrary frequency. For any
$\omega_{k} = \omega_{0} + 2\pi k/T$ with $k$ an integer we have
that $\hat{\ba}(e^{j\omega_k T})= \hat{\ba}(e^{j\omega_0 T})$.
Therefore, by evaluating $\hat{y}$ and $\hat{\bphil}$ at
frequencies $-\pi/T_{c} \leq \omega_{k} \leq \pi/T_c$, we can use
(\ref{eq:seq}) to generate more equations. To this end, let $L$ be
the largest integer for which $L \leq T/T_{c}$. Assume first that
$L = 2 L_{0} + 1$ for some integer $L_{0}$, so that $L$ is odd.
 We then generate the equations
\begin{equation*}
    \hat{y}_{k}(\omega) := \hat{y}(\omega- k\, \tfrac{2\pi}{T})=\sum_{n=1}^{N} \hat{\phi}_{n}(\omega - k\, \tfrac{2\pi}{T})\, \hat{a}_{n}(\omega - k\, \tfrac{2\pi}{T})
\end{equation*}
for $-L_{0} \leq k \leq L_{0}$ and for $\omega\in [-\pi/T,\pi/T]$.
Since by our assumption $\pi/T_c \geq L\, \pi/T$, all the
observations $\hat{y}_{k}(\omega) = \hat{y}(\omega - 2 k\, \pi/T)$
are in the passband regime of the LPF. The above set of $L$
equations may be written as
\begin{equation}
\label{eq:meq}
    \hat{\mathbf{y}}(\omega) = \bphi_{L}(\omega)\,\hat{\ba}(e^{j\omega T})\,,
    \quad \omega\in [-\pi/T,\pi/T],
\end{equation}
where $\hat{\mathbf{y}}(\omega) = [\hat{y}_{-L_{0}}(\omega),
\dots, 0, \dots, \hat{y}_{L_{0}}(\omega)]^{T}$ is a length $L$
vector containing all the different observations $\hat{y}_{k}$ of
the output $\hat{y}$, and $\bphi_{L}(\omega)$ is the $L \times N$
matrix given by \eqref{eq:phim}. In the case where $L = 2 L_{0}$
is an even number\footnote{In subsequent sections, we will only
discuss the case where $L$ is odd. The necessary changes for the
case of $L$ being even are obvious.}, we generate additional
equations by
\begin{equation}
\label{eq:yk}
    \hat{y}_{k}(\omega)
    := \sum_{n=1}^{N} \hat{\phi}_{n}(\omega - [2k+1]\, \tfrac{\pi}{T})\, \hat{a}_{n}(\omega - [2k+1]\, \tfrac{\pi}{T})
\end{equation}
for $-L_{0} \leq k  \leq L_{0}-1$. Here again all the observations
in (\ref{eq:yk}) are in the passband regime of the LPF. Therefore,
(\ref{eq:yk}) can be written as in \eqref{eq:meq} where
$\bphi_{L}(\omega)$ is now given by \eqref{eq:phim2}, and the
definition of $\hat{\ba}$ is changed accordingly.

If the matrix $\bphi_{L}(\omega)$ satisfies the sufficient
conditions of Proposition~\ref{lem:SuffCond}, then the unknown
vector $\hat{\ba}\ejt$ can be recovered from (\ref{eq:meq}) by
solving the linear set of equations for all $\omega
\in[-\pi/T,\pi/T]$. In particular, there exists a left inverse
$\bbg(\omega)$ of $\bphi_{L}(\omega)$ such that
$\hat{\ba}(e^{j\omega T}) = \bbg(\omega)\,
\hat{\mathbf{y}}(\omega)$. Finally, the desired sequences
$\{a_{n}[k]\}_{k\in\ZZ}$ are the Fourier coefficients of the
$2\pi/T$ periodic functions $\hat{a}_{n}$.

\section{Preprocessing With Filters}
\label{sec:Filter}

When $\bphi_{L}(\omega)$ does not has full column rank for all
$\omega\in [-\pi/T,\pi/T]$ and if the bandwidth of the LPF can not
be increased, an interesting question is whether we can
pre-process $x$ before lowpass filtering in order to ensure that
it can be recovered from the LPF output. In this and in the next
section we consider two types of pre-processing: using a bank of
filters, and using a bank of mixers (modulators), respectively.

\begin{figure}[h]
\begin{center}
\begin{picture}(40,20)(0,0)

\put(-3,-3.5){
\put(9,6.5){{\ar{\filter(8,5){$g_N(t)$}{\ar{\filter(8,5){$$}{\ar{\et{$y_N(t)$}}}}}}}}

\put(.1,.7){
    \put(26,5.5){\line(1,0){6}}
    \put(27,5.5){\line(0,1){1.7}}
    \put(29,5.5){\line(0,1){2.3}}
    \put(31,5.5){\line(0,1){1.7}}
    \put(27,7.2){\line(1,0){4}}
    \put(27.0,4.5){\makebox(0,0){\footnotesize $-\pi/T_c$}}
    \put(31.4,4.5){\makebox(0,0){\footnotesize $\pi/T_c$}}
}

\put(0,14){

\put(9,6.5){{\ar{\filter(8,5){$g_1(t)$}{\ar{\filter(8,5){$$}{\ar{\et{$y_1(t)$}}}}}}}}

\put(.1,.7){
    \put(26,5.5){\line(1,0){6}}
    \put(27,5.5){\line(0,1){1.7}}
    \put(29,5.5){\line(0,1){2.3}}
    \put(31,5.5){\line(0,1){1.7}}
    \put(27,7.2){\line(1,0){4}}
    \put(27.0,4.5){\makebox(0,0){\footnotesize $-\pi/T_c$}}
    \put(31.4,4.5){\makebox(0,0){\footnotesize $\pi/T_c$}}
}

}

\put(5.5,13.5){\st{$x(t)$}{\hspace*{-0.05in}\sar}}
\put(9,6.5){\line(0,1){14}}

\put(17,13){\Large $\vdots$} \put(29,13){\Large $\vdots$} }
\end{picture}
\end{center}
\caption{Preprocessing of $x(t)$ by a bank of $N$ LTI filters.}
\label{fig:pref}
\end{figure}
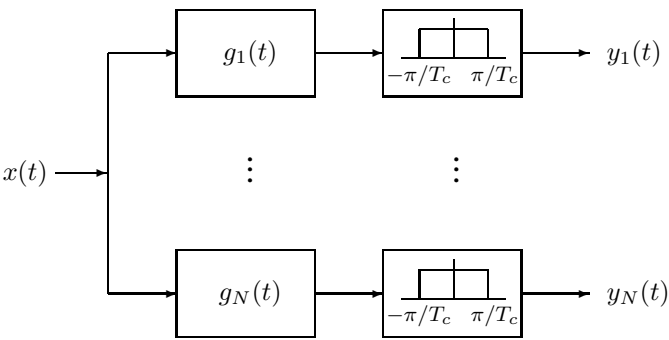

Suppose we allow pre-processing of $x$ with a set of $N$ filters,
as in Fig.~\ref{fig:pref}. The question is whether we can choose the
filters $g_{n}$ in the figure so that $x$ can be recovered
from the outputs $y_{n}$ of each of the branches under more
mild conditions than those developed in Section~\ref{sec:RecovCond}.

Let $\hat{\by},\hat{\bg}$ be the length-$N$ vectors with
$n$th elements given by $\hat{y}_{n},\hat{g}_{n}$.
Then we can immediately verify that
\begin{equation}
\label{eq:seqf}
    \hat{\by}(\omega) =
    \hat{\bg}(\omega)\, \hat{\bphil}^T(\omega)\, \hat{\ba}\ejt,
    \quad \omega \in [-\tfrac{\pi}{T_c},\tfrac{\pi}{T_c}].
\end{equation}
Clearly, $\hat{\ba}$ cannot be recovered from this set of equations
as all the equations are linearly dependent (they are all
multiples of each other). Thus, although we have $N$ equations,
only one of them provides independent information on $\hat{\ba}$. We
can, as before, use the periodicity of $\hat{\ba}$ if $T_c$ is small
enough. Following the same reasoning as in Section~\ref{sec:RecovAlg}, assuming that
$\pi/T_{c} \geq L\, \pi/T$, we can create $L-1$ new measurements using the same
unknowns $\hat{\ba}$ by considering $\hat{\by}(\omega)$ for different
frequencies $\omega + k 2\pi /T$. In this case though it is
obvious that the pre-filtering does not help, since only one
equation can be used from the set of $N$ equations (\ref{eq:seqf})
for each frequency. In other words, all the branches in
Fig.~\ref{fig:pref} provide the same information. The resulting
equation is the same as in the previous section up to
multiplication by $\hat{g}_{n}$ for one index $1 \leq n \leq N$.
Therefore, the recovery conditions reduce to the same ones as
before, and having $N$ branches does not improve our ability to
recover $x$.

\section{Preprocessing With Mixers}
\label{sec:Mixers}

We now consider a different approach, which as we will see leads
to greater benefit. In this strategy, instead of using filters in
each branch, we use periodic mixing functions $p_{n}$. Each
sequence is assumed to be periodic with period equal
to\footnote{Note, that we can also choose $T_p=T/r$ for an integer
$r$. However, for simplicity we restrict attention to the case
$r=1$.} $T_{p} = T$. By choosing the mixing functions
appropriately, we can increase the class of functions that can be
recovered from the lowpass filtered outputs.

\subsection{Single Channel}

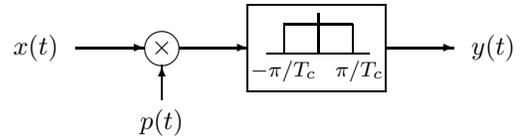
\begin{figure}[h]
\begin{center}
\begin{picture}(32,8)(0,0)

\put(4,5){\st{$x(t)$}{\ar{\rmult{\ar{\filter(8,5){$$}{\ar{\et{$y(t)$}}}}}}}}

\put(-10,-1.5){

\put(.1,.7){
    \put(26,5.5){\line(1,0){6}}
    \put(27,5.5){\line(0,1){1.7}}
    \put(29,5.5){\line(0,1){2.3}}
    \put(31,5.5){\line(0,1){1.7}}
    \put(27,7.2){\line(1,0){4}}
    \put(27.0,4.5){\makebox(0,0){\footnotesize $-\pi/T_c$}}
    \put(31.4,4.5){\makebox(0,0){\footnotesize $\pi/T_c$}}
}

}

\put(10,.7){\makebox(0,0){$p(t)$}} \put(10,2){\ssuar}

\end{picture}
\end{center}
\caption{Mixing prior to lowpass filtering of $x(t)$.}
\label{fig:mf}
\end{figure}

Let us begin with the case of a single mixing function, as in
Fig.~\ref{fig:mf}.
Since $p$ is assumed to be periodic with period $T$, it can be written as a Fourier series
\begin{equation}
\label{eq:ptSeries}
    p(t) = \sum_{k\in\bbZ} b_{k}\, \expp{j 2\pi kt/T}
\end{equation}
where
\begin{equation}
\label{eq:FourierCoeff}
    b_{k}=\frac{1}{T}\int_{-T/2}^{T/2} p(t)\,\expp{-j 2\pi kt/T}d t\, ,
    \quad k\in\bbZ
\end{equation}
are the Fourier coefficients of $p$.
The sum \eqref{eq:ptSeries} is assumed to converge in $L^{2}$ which implies that the sequence $\{b_{k}\}_{k\in\bbZ}$ is an element of $\ell^{2}$.
The output $y = P_{\pi/T_{c}}(p \, x)$ of the LPF is then given in the frequency domain by
\begin{equation}
\label{eq:ym}
    \hat{y}(\omega)
    = \sum_{k \in \ZZ} b_k\, \hat{x}(\omega - k \tfrac{2\pi}{T}),
    \quad \omega \in [-\tfrac{\pi}{T_{c}},\tfrac{\pi}{T_{c}}].
\end{equation}
Using (\ref{eq:xeq}) and the fact that $\hat{a}_{n}\ejt$ is
$2\pi/T$-periodic, (\ref{eq:ym}) can be written as
\begin{equation}
\label{eq:ymp}
    \hat{y}(\omega)
    = \sum_{n=1}^N  \hat{a}_{n}\ejt
    \sum_{k \in \ZZ} b_{k}\, \hat{\phi}_{n}(\omega - k \tfrac{2\pi}{T}),
\end{equation}
for $\omega \in [-\pi/T_{c}, \pi/T_{c}]$. Defining
\begin{equation}
\label{eq:zdef}
    \hat{\gamma}_{n}(\omega)
    : = \sum_{k \in \ZZ} b_{k} \hat{\phi}_{n}(\omega - k \tfrac{2\pi}{T}),
    \quad 1 \leq n \leq N
\end{equation}
and denoting by $\hat{\bpgamma}$ the vector whose $n$th element is
$\hat{\gamma}_{n}$, we can express \eqref{eq:ymp} as
\begin{equation}
\label{eq:yz}
    \hat{y}(\omega)
    = \hat{\bpgamma}^T(\omega)\,\hat{\ba}\ejt,
    \quad \omega \in[-\tfrac{\pi}{T_c},\tfrac{\pi}{T_c}].
\end{equation}

Equation (\ref{eq:yz}) is similar to (\ref{eq:seq}) with
$\hat{\bpgamma}$ replacing $\hat{\bphil}$. Therefore, as in the
case in which no pre-processing took place (cf.
Section~\ref{sec:RecovAlg}), we can create $L - 1$ additional
equations by evaluating $\hat{y}(\omega)$ at frequencies $\omega +
2 k\, \pi/T$ as long as $\pi/T_{c} \geq L\, \pi/T$. This yields
the system of equations
\begin{equation}
\label{equ:EQU_withMixing}
    \hat{\mathbf{y}}(\omega)
    = \bgamma_{L}(\omega)\,\hat{\ba}(e^{j\omega T})\,,
    \quad \omega\in [-\tfrac{\pi}{T},\tfrac{\pi}{T}],
\end{equation}
where $\hat{\mathbf{y}}$ and $\hat{\ba}$ are defined as in
\eqref{eq:phim} and
\begin{eqnarray*}
    \bgamma_{L}(\omega) =
    \left[ \begin{array}{ccc}
    \hat{\gamma}_1(\omega + L_{0} \tfrac{2\pi}{T})  & \ldots & \hat{\gamma}_N(\omega + L_{0} \tfrac{2\pi}{T})\\
    \vdots &&\vdots \\
    \hat{\gamma}_1(\omega) & \ldots & \hat{\gamma}_N(\omega) \\
    \vdots &&\vdots \\
    \hat{\gamma}_1(\omega - L_{0} \tfrac{2\pi}{T} ) & \ldots & \hat{\gamma}_N(\omega - L_{0} \tfrac{2\pi}{T} )  \\
    \end{array} \right].
\end{eqnarray*}
Consequently, we can recover $\hat{\ba}$ from the given measurements as long as the matrix $\bgamma_{L}(\omega)$
has full column rank for all $\omega \in [-\pi/T,\pi/T]$. To this end it is necessary that $\pi/T_{c} \geq N \pi/T$, i.e. that $L\geq N$.

Due to the mixing of the signal, the coefficient matrix $\bphi_{L}(\omega)$ in \eqref{eq:meq} is changed to $\bgamma_{L}(\omega)$ in \eqref{equ:EQU_withMixing}. This new coefficient matrix is constructed out of the "new generators" $\{\gamma_{n}\}^{N}_{n=1}$ in exactly the same way as $\bphi_{L}(\omega)$ is constructed from the original generators $\{\phi_{n}\}^{N}_{n=1}$.
Equation \eqref{eq:zdef} shows that the Fourier transform $\hat{\gamma}_{n}$ of each new generator lies in a shift invariant space
\begin{eqnarray*}
    \mathcal{S}_{\frac{2\pi}{T}}(\hat{\phi}_{n})
    = \cls\{ S_{k \frac{2\pi}{T}}\, \hat{\phi}_{n} : k\in\bbZ \}
\end{eqnarray*}
spanned by shifts of $\hat{\phi}_{n}$. The coefficients $\{b_{k}\}_{k\in\bbZ}$ of the mixing sequence are then the "coordinates" of $\hat{\gamma}_{n}$ in  $\mathcal{S}_{\frac{2\pi}{T}}(\hat{\phi}_{n})$.

We now want to show that the condition of invertibility of $\bgamma_{L}(\omega)$ is in general easier to
satisfy then the analogous condition on the matrix $\bphi_{L}(\omega)$
of (\ref{eq:phim}). To this end, we write $\bgamma_{L}(\omega)$ as
\begin{equation}
\label{eq:zd}
    \bgamma_{L}(\omega) = \bbb_{L}\,\bphi(\omega),
\end{equation}
where $\bphi(\omega)$ denotes the matrix consisting of $N$ columns
and infinitely many rows $\hat{\bphil}^T(\omega + k\, 2\pi/T)$
with $k \in \ZZ$. Note that $\bphi(\omega)$ has the form
\eqref{eq:phim} with $L \to \infty$, i.e. $\bphi(\omega) =
\bphi_{\infty}(\omega)$. The matrix $\bbb_{L}$ with $L = 2 L_{0} +
1$ rows and infinite columns contains the Fourier coefficients
$\{b_{k}\}_{k\in\bbZ}$ of the mixing sequence \eqref{eq:ptSeries}
and is given by
\begin{eqnarray}
\label{eq:bB_1}
    \bbb_{L} =
    \left[ \begin{array}{ccccc}
    \dots  & b_{L_{0}-1}  & b_{L_{0}}  & b_{L_{0}+1}  & \dots \\
    \      &        & \vdots &        &  \\
    \dots  & b_{0}  & b_{1}  & b_{2}  & \dots \\
    \dots  & b_{-1} & b_{0}  & b_{1}  & \dots \\
    \dots  & b_{-2} & b_{-1} & b_{0}  & \dots \\
    \      &        & \vdots &        & \\
    \dots  & b_{-L_{0}-1} & b_{-L_{0}} & b_{-L_{0}+1} & \dots \\
    \end{array} \right].
\end{eqnarray}
Representation \eqref{eq:zd} follows immediately from the relation
$\hat{\gamma}_{n}(\omega - \ell\, 2\pi/T) = \sum_{k\in\bbZ}
b_{k-\ell} \hat{\phi}_{n}(\omega - k \tfrac{2\pi}{T})$ for the
entries of the matrix $\bgamma_{L}(\omega)$.

The Grammian $\bbm_{\phi}(\omega)$ of the generators $\phi$,
defined in \eqref{eq:Mf}, can be written as $\bbm_{\phi}(\omega) =
\bphi^{*}(\omega)\bphi(\omega)$. Therefore, under our assumption
(\ref{eq:rc}) on the generators, $\bphi(\omega)$ has full column
rank for all $\omega \in[-\pi/T,\pi/T]$. The question then is
whether we can choose the sequence $\{b_{k}\}_{k\in\bbZ} \in
\ell^{2}$, and consequently the function $p$, so that
$\bbb_{L}\,\bphi(\omega)$ has full-column rank i.e. such that the
matrix $\bgamma^{*}_{L}(\omega)\bgamma_{L}(\omega) =
\bphi^{*}(\omega)\bbb^{*}_{L}\bbb_{L}\bphi(\omega)$ is invertible
for all $\omega \in [-\pi/T,\pi/T]$.

If we choose the mixing sequence $p(t)\equiv 1$ then $b_{0} = 1$
and $b_{k} = 0$ for all $k\neq 0$. Consequently
$\bbb_{L}\bphi(\omega)$ is comprised of the first $L$ rows of
$\bphi(\omega)$, so that $\bgamma_{L}(\omega) =
\bphi_{L}(\omega)$. However, by allowing for general sequences
$\{b_{k}\}_{k\in\bbZ}$, we have more freedom in choosing
$\bbb_{L}$ such that the product $\bbb_{L}\, \bphi(\omega)$ may
have full column-rank, even if $\bphi_{L}(\omega)$ does not.

We next give a simple example which demonstrates that
pre-processing by an appropriate mixing function can enable the
recovery of the signal.

\begin{example}
We continue Example~\ref{ex:sinc_N1} with the single generator
$\phi_{1}$ given by \eqref{equ:Sinc_Example}. Here we assume that
the parameter $D$ satisfies the relation $1 < D/T < 3/2$ and that
the cutoff frequency of the lowpass filter is $\pi/T_{c} = \pi/T$.
In this case, recovery of $x$ from its lowpass component $y =
P_{\pi/T_{c}}x$ is not possible, as discussed in
Example~\ref{ex:sinc_N1}. However, we will show that there exist
mixing functions $p$ so that $x$ can be recovered from $y =
P_{\pi/T_{c}}(p x)$.

One possible mixing function is
\begin{equation*}
    p(t) = 1 + 2\, \sin(2\pi t/T)
\end{equation*}
whose Fourier coefficients \eqref{eq:FourierCoeff} are given by
$b_{-1} = -j$, $b_{0} = 1$, $b_{1} = j$, and $b_{k} = 0$ for all
$|k|\geq 2$. With this choice, the "new generator" \eqref{eq:zdef}
becomes
\begin{multline*}
    \hat{\gamma}_{1}(\omega)
    = \frac{\sin(\omega D)}{\omega D}\\
    + j \left[ \frac{\sin(\omega D - 2\pi D/T)}{\omega D - 2\pi D/T}
    - \frac{\sin(\omega D + 2\pi D/T)}{\omega D + 2\pi D/T} \right].
\end{multline*}
Since $\pi/T_{c} = \pi/T$, the matrix $\bgamma_{L}(\omega)$
reduces to the scalar $\hat{\gamma}_{1}(\omega)$ and we have to
show that $0 < |\hat{\gamma}_{1}(\omega)| < \infty$ for all
$\omega \in [-\pi/T,\pi/T]$. The upper bound is trivial; for the
lower bound, it is sufficient to show that the real and imaginary
part of $\hat{\gamma}_{1}$ have no common zero in
$[-\pi/T,\pi/T]$. This fact is easily verified by noticing that
the only zeros of the real part of $\hat{\gamma}_{1}(\omega)$ are
at $\omega_{1} = \pi/D$ and $\omega_{2} = - \pi/D$. Evaluating the
imaginary part $\Im\{\hat{\gamma}_{1}\}$ of $\hat{\gamma}_{1}$ at these zeros gives
\begin{eqnarray*}
    \left| \Im\{\hat{\gamma}_{1}(\omega_{1,2})\} \right|
    = \frac{1}{2\pi} \frac{|\sin(2\pi D/T)|}{(D/T)^{2} - 1/4}
\end{eqnarray*}
which is non-zero under the assumption made on $D/T$.
\end{example}

The general question whether for a given set $\phi =
\{\phi_{1},\dots,\phi_{N}\}$ of generators there exists a matrix
$\bbb_{L}$ such that \eqref{eq:zd} is invertible for all
$\omega\in [-\pi/T,\pi/T]$, or under what conditions on the
generators $\phi$ such a matrix can be found seems to be an open
and non-trivial question.  The major difficulty is that according
to \eqref{eq:zd}, we look for a constant (independent of $\omega$)
matrix $\bbb_{L}$ such that $\bbb_{L}\, \bphi(\omega)$ has full
column rank for all $\omega \in [-\pi/T,\pi/T]$. Moreover, the
matrix $\bbb_{L}$ has to be of the particular form \eqref{eq:bB_1}
with a sequence $\{b_{k}\}_{k\in\bbZ} \in \ell^{2}$.

The next example characterizes a class of generators for which a
simple (trivial) mixing sequence always exist.

\begin{example}[generators with compact support]
\label{ex:CompactSupport} \sloppy We consider the case of a single
generator ($N=1$) and assume that $\pi/T_{c} = \pi/T$, i.e. $L = N
= 1$. Our problem then reduces to finding a function
$\hat{\gamma}_{1} \in
\mathcal{S}_{\frac{2\pi}{T}}(\hat{\phi}_{1})$ such that
$\hat{\gamma}_{1}(\omega) \neq 0$ for all $\omega \in
[-\pi/T,\pi/T]$.

We treat the special case of a generator $\phi_{1}$ with finite
support of the form $[-D,D]$ for some $D\in\bbR$, i.e. we assume
that $\phi_{1}(t) = 0$ for all $t \notin [-D,D]$. This means that
its Fourier transform $\hat{\phi}_{1}$ is an element of the
Paley-Wiener space $\widehat{PW}(D)$ and so are all linear
combinations of the shifts $S_{2\pi/T}\hat{\phi}_{1}$. It follows
that $\mathcal{S}(\hat{\phi}_{1}) \subset \widehat{PW}(D)$.

Let now $\hat{\gamma} \in \mathcal{S}(\hat{\phi}_{1})$ be arbitrary and let $\{\omega_{k}\}_{k\in\ZZ}$ be the ordered sequence of real zeros of $\hat{\gamma}$ with $\omega_{n} \leq \omega_{n+1}$. Then a theorem of Walker \cite{Walker_1991} states that
\begin{equation*}
    \sup_{n\in\bbZ}|\omega_{n+1} - \omega_{n}| > \pi/D.
\end{equation*}
Thus there exists at least one interval of the real line of length
$\pi/D$ such that $\hat{\gamma}$ has no zeros in this interval.
Consequently, if $\pi/D > 4\pi/T$ then there always exists a
$k_{0}\in\bbZ$ such that
\begin{equation}
    \hat{\gamma}(\omega - k_{0}\, \tfrac{2\pi}{T}) \neq 0
    \quad\text{for all}\ \omega\in [-\tfrac{\pi}{T},\tfrac{\pi}{T}].
\end{equation}
This holds in particular for the generator $\hat{\phi}_{1}$ itself.

We conclude that if the support of the generator $\phi_{1}$
satisfies $\supp(\phi_{1}) < T/4$, then there always exists a
$k_{0} \in \bbZ$ such that $\hat{\gamma}_{1}(\omega) =
\hat{\phi}_{1}(\omega - k_{0}\, 2\pi/T) \neq 0$ for all $\omega
\in [-\pi/T,\pi/T]$.  The corresponding mixing sequence is given
by $b_{k_{0}} = 1$ and $b_{k} = 0$ for all $k \neq k_{0}$.
\end{example}

\subsection{Multiple Channels}

In the single channel case, it was necessary that the cutoff
frequency $\pi/T_{c}$ of the LPF is at least $N$ times larger than
the bandwidth of the desired signal $\hat{\ba}$ in order to be
able to recover the signal. We will now show that using several
channels can reduce the cutoff frequency $\pi/T_{c}$ of the filter
in each channel, from which we can still recover the original
signal $x$.

\begin{figure}[h]
\begin{center}
\begin{picture}(34,23)(0,0)

\put(-2,-1.5){ \put(9,6.5){{\ar{\rmult{\ar{\filter(8,5){$$}{\ar{\et{$y_L(t)$}}}}}}}}

\put(14,2.5){\makebox(0,0){$p_{L}(t)$}} \put(14,3.5){\ssuar}

\put(-6,.7){
    \put(26,5.5){\line(1,0){6}}
    \put(27,5.5){\line(0,1){1.7}}
    \put(29,5.5){\line(0,1){2.3}}
    \put(31,5.5){\line(0,1){1.7}}
    \put(27,7.2){\line(1,0){4}}
    \put(27.0,4.5){\makebox(0,0){\footnotesize $-\pi/T_c$}}
    \put(31.4,4.5){\makebox(0,0){\footnotesize $\pi/T_c$}}
}

 \put(0,14){

 \put(9,6.5){{\ar{\rmult{\ar{\filter(8,5){$$}{\ar{\et{$y_1(t)$}}}}}}}}

\put(14,2.5){\makebox(0,0){$p_{1}(t)$}} \put(14,3.5){\ssuar}

\put(-6,.7){
    \put(26,5.5){\line(1,0){6}}
    \put(27,5.5){\line(0,1){1.7}}
    \put(29,5.5){\line(0,1){2.3}}
    \put(31,5.5){\line(0,1){1.7}}
    \put(27,7.2){\line(1,0){4}}
    \put(27.0,4.5){\makebox(0,0){\footnotesize $-\pi/T_c$}}
    \put(31.4,4.5){\makebox(0,0){\footnotesize $\pi/T_c$}}
}

}

\put(5.5,13.5){\st{$x(t)$}{\hspace*{-0.05in}\sar}}
\put(9,6.5){\line(0,1){14}}

\put(14,12){\Large $\vdots$} \put(23,12){\Large $\vdots$} }
\end{picture}
\end{center}
\caption{Bank of mixing functions.} \label{fig:prem}
\end{figure}
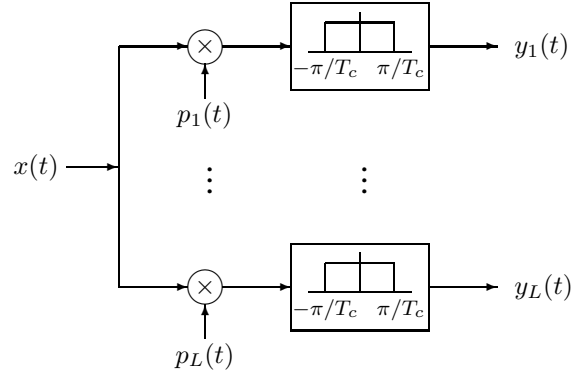

Suppose that we have $L \geq N$ channels, where each channel uses
a different mixing sequence, as in Fig.~\ref{fig:prem}.
Since $L \geq N$, we expect to be able to reduce the cutoff in
each channel. We therefore consider the case in which $T_c=T$. The
output $y_{\ell} = P_{\pi/T}(p_{\ell}\, x)$ of the $\ell$th
channel in the frequency domain is then equal to
\begin{equation*}
    \hat{y}_{\ell}(\omega)
    = \hat{\bpgamma}_{\ell}^T(\omega) \hat{\ba}\ejt,
    \quad \omega \in [-\tfrac{\pi}{T}, \tfrac{\pi}{T}]
\end{equation*}
where $\hat{\bpgamma}_{\ell}^T(\omega)$ is the vector with $n$th
element
\begin{equation*}
    [\hat{\bpgamma}_{\ell}(\omega)]_{n}
    =  \hat{\gamma}^{\ell}_{n}(\omega)
    := \sum_{k \in \ZZ} b_k^{\ell} \hat{\phi}_{i}(\omega + k\tfrac{2\pi}{T}),
\end{equation*}
and $\{b_{k}^{\ell}\}_{k\in\bbZ}$ are the Fourier coefficients
associated with the $\ell$th sequence $p_{\ell}$. Defining by
$\hat{\by}(\omega)$ the vector with $\ell$th element
$\hat{y}_{\ell}(\omega)$ we conclude that
\begin{equation*}
    \hat{\by}(\omega)
    = \bgamma_{L}(\omega)\,\hat{\ba}\ejt,
    \quad \omega \in [-\tfrac{\pi}{T},\tfrac{\pi}{T}]
\end{equation*}
where $\bgamma_{L}(\omega)$ is the matrix whose entry in the
$\ell$th row and $n$th column is $[\bgamma(\omega)]_{\ell,n} =
\hat{\gamma}^{\ell}_{n}(\omega)$. Now, all we need is to choose
the $L$ sequences $\{b_{k}^\ell\}_{k\in\bbZ} \in \ell^{2}$ such
that $\bgamma_{L}(\omega)$ has full column rank. More
specifically, as before we can write
\begin{equation}
\label{eq:DPHi}
    \bgamma_{L}(\omega) = \bbb_{L}\bphi(\omega),
\end{equation}
where $\bbb_{L}$ is a matrix with $L$ rows and infinitely many columns whose $\ell$th row is given by the coefficient sequence $\{b_{k}^\ell\}_{k\in\bbZ}$, i.e.
\begin{eqnarray*}
    \bbb_{L} =
    \left[ \begin{array}{ccccccc}
    \dots  & b^{1}_{-2} & b^{1}_{-1} & b^{1}_{0}  & b^{1}_{1}  & b^{1}_{2} & \dots \\
    \dots  & b^{2}_{-2} & b^{2}_{-1} & b^{2}_{0}  & b^{2}_{1}  & b^{2}_{2} & \dots \\
    \      &        &        & \vdots &        &       & \\
    \dots  & b^{L}_{-2} & b^{L}_{-1} & b^{L}_{0}  & b^{L}_{1}  & b^{L}_{2} & \dots \\
    \end{array} \right].
\end{eqnarray*}
By our assumption $\bphi(\omega)$ has full column rank and so it remains to choose $\bbb_{L}$ such that $\bgamma_{L}(\omega)$ is invertible for every $\omega \in [-\pi/T,\pi/T]$.

It should be noted that we used the same notation as in the
previous subsection although the definition of the particular
matrices and vectors differ slightly in both cases. Nevertheless,
the formal approach is very similar. In the previous subsection,
we observed the output signal in different frequency channels $1
\leq \ell \leq L$ whereas in this subsection the channels $1 \leq
\ell \leq L$ are characterized by different mixing
sequences\footnote{In the first case we perform "frequency
multiplexing" whereas the second case resembles "code
multiplexing".}.

As in the previous subsection, the general question whether for a
given system $\phi = \{\phi_{1},\dots,\phi_{N}\}$ of generators
there always exists an appropriate system of mixing sequences $p =
\{p_{1}, \dots, p_{L}\}$ such that $\bgamma_{L}(\omega)$ has full
column rank for all frequencies $\omega$ seems to be non-trivial.
The formal difficulty lies in the fact that we look for a constant
(independent of $\omega$) matrix $\bbb_{L}$ such that
\eqref{eq:DPHi} has full column rank for each $\omega \in
[-\pi/T,\pi/T]$. However, compared with the previous section,
where only one mixing sequence was applied, the problem of finding
an appropriate matrix $\bbb_{L}$ becomes simpler: In the former
case $\bbb_{L}$ has to have the special (diagonal) form
\eqref{eq:bB_1}, whereas here its entries can be chosen (almost)
arbitrarily. The sequences $\{b^{\ell}_{k}\}_{k\in\bbZ}$ only have
to be in $\ell^{2}$.

A special choice of periodic functions that are easy to implement
in practice are binary sequences. This example was studied in
\cite{ME09} in the context of sparse multiband sampling. More
specifically, $p_{\ell}$, $1 \leq \ell \leq L$ are chosen to
attain the values $\pm 1$ over intervals of length $T/M$ where $M$
is a given integer.  Formally,
\begin{equation}
\label{eq:ptsl}
    p_\ell(t) = \alpha^\ell_{n}\,,
    \ \ n\tfrac{T}{M} \leq t < (n+1)\tfrac{T}{M}\,,
    \ \ 0\leq n\leq M-1
\end{equation}
with $\alpha^\ell_{n}\in\{+1,-1\}$, and $p_\ell(t+k T) =
p_\ell(t)$ for every $k \in\bbZ$. In this case, we have
\begin{align*}
    b^{\ell}_{k} & = \frac{1}{T}\int_0^{T}p_{\ell}(t)\, \expp{-j\frac{2\pi}{T}kt}d t \nonumber\\
                & = \frac{1}{T} \int_0^{T/M} \sum_{n=0}^{M-1} \alpha^{\ell}_{n} \expp{-j\frac{2\pi}{T}k(t+n\frac{T}{M})}dt\nonumber\\
                & = \frac{1}{T} \sum_{n=0}^{M-1} \alpha^{\ell}_{n}
                    \expp{-j\frac{2\pi}{M}nk}\int_0^{T/M}\expp{-j\frac{2\pi}{T}kt}dt.
\end{align*}
Evaluating the integral gives
\begin{align*}
    b_0^\ell = \frac{1}{M} \hat{\alpha}_0^\ell
    \quad\text{and}\quad
    b_k^\ell =\frac{1-e^{-j\omega_0k}}{j 2\pi k} \hat{\alpha}_k^\ell,
    \quad k\neq 0
\end{align*}
where $\omega_0 = 2\pi/M$, and $\{\hat{\alpha}_k^\ell\}_{k\in\bbZ}$ denotes the discrete Fourier transform (DFT) of the sequence $\{\alpha^\ell_{n}\}^{M-1}_{n=0}$. Note that $\{\hat{\alpha}_k\}_{k\in\bbZ}$ is $M$-periodic so that $\hat{\alpha}_k=\hat{\alpha}_{k+M}$.

With these mixing sequences, the infinite matrix $\bbb_{L}$ can be written as
\begin{equation}
\label{eq:QFW}
    \bbb_{L} = \bbq\bbf^*\bbw,
\end{equation}
where $\bbq$ is a matrix with $M$ columns and $L$ rows, whose
$\ell$th row is given by the sequence
$\{\hat{\alpha}^\ell_n\}^{M-1}_{n=0}$, $\bbf$ is the $M \times M$
Fourier matrix, and $\bbw$ is a matrix with $M$ rows and
infinitely many columns consisting of block diagonal matrices of
size $M \times M$ whose diagonal values are given by the sequence
$\{w_{k}\}_{k\in\bbZ}$ defined by $w_{0} = 1/M$ and $w_{k} =
\frac{1-e^{-j\omega_0k}}{j 2\pi k}$ for $k\neq 0$. Applying these
binary mixing sequences, the problem is now to find a finite $L
\times N$ matrix $\bbq$ with values in $\{+1,-1\}$ such that
$\bbq\bbf^*\bbw\, \bphi(\omega)$ has full column rank for every
$\omega\in [-\pi/T,\pi/T]$.

The next example shows how to select $\bbq$ in the case of
bandlimited generators.

\begin{example}[bandlimited generators]
\label{ex:BandlimitedGenerators} We consider the case where each
generator $\phi_{n}$ is bandlimited to the interval $[-K_{0}\,
\pi/T, K_{0}\, \pi/T]$ for some $K_{0}\in\mathbb{N}$, and $N = 2
K_{0} + 1$. In this case, $\mathbf{\Phi}(\omega) =
\mathbf{\Phi}_{N}(\omega)$ is essentially an $N \times N$ matrix
(all other entries are identically zero). This matrix is
invertible for every $\omega \in [-\pi/T,\pi/T]$ according to
assumption \eqref{eq:rc}.

We now apply $L = N$ different mixing sequences
$\{p_{\ell}\}^{L}_{\ell=1}$ having the special structure
\eqref{eq:ptsl}, and choose $M = N$. According to \eqref{eq:DPHi}
and \eqref{eq:QFW} the matrix $\bgamma_{L}(\omega)$ then becomes
\begin{equation}
\label{equ:GammaL_Exa}
    \bgamma_{L}(\omega)
  = \mathbf{Q}\, \mathbf{F}^{*}\, \mathbf{W}\,
  \mathbf{\Phi}(\omega),
\end{equation}
where $\mathbf{Q}\, \mathbf{F}^{*}$ and $\mathbf{W}\,
\mathbf{\Phi}(\omega)$ are matrices of size $N \times N$. The
matrix $\mathbf{W}\, \mathbf{\Phi}(\omega)$ may be considered as
the product of the invertible $N\times N$ matrix
$\mathbf{\Phi}(\omega) = \mathbf{\Phi}_{N}(\omega)$ with an
$N\times N$ diagonal matrix consisting of the central diagonal
matrix of $\mathbf{W}$, i.e.
\begin{equation*}
    \mathbf{W}\, \mathbf{\Phi}(\omega)
    = \diag(w_0, \ldots,w_{N-1})\, \bphi_{N}(\omega).
\end{equation*}
Since this diagonal matrix is invertible also $\mathbf{W}\,
\mathbf{\Phi}_{N}(\omega)$ is invertible for every $\omega \in
[-\pi/T,\pi/T]$. Therefore, using the fact that the Fourier matrix
$\mathbf{F}$ is invertible, $\bgamma_{L}(\omega)$ is invertible
for each $\omega \in [-\pi/T,\pi/T]$ if the values
$\{\alpha^{\ell}_{n}\}^{N}_{n=1}$ of the mixing sequences
$p_{\ell}$ are chosen such that $\mathbf{Q}$ is invertible. This
can be achieved by choosing $\mathbf{Q}$ as a Hadamard matrix of
order $N$. It is known that Hadamard matrices exists at least for
all orders up to $667$ \cite{KT-R05_Hadamard428}.
\end{example}

In the previous example, $\mathbf{\Phi}_{N}(\omega)$ was an $N
\times N$ invertible matrix for all $\omega \in [-\pi/T,\pi/T]$.
According to Proposition~\ref{lem:SuffCond} recovery of the signal
$x$ is therefore possible if the bandwidth of the LPF is larger
than $N \pi/T$. However, the example shows that pre-processing of
$x$ by applying the binary sequences in $L = N$ channels allows
recovery of the signal already from its signal components in the
frequency range $[-\pi/T,\pi/T]$.

For simplicity of the exposition, we assumed throughout this
subsection that the bandwidth $2\pi/T_{c}$ of the lowpass filter
is equal to the signal bandwidth $2 \pi/T$ and that the number of
channels $L$ is at least equal to the number of generators $N$.
However, it is clear from the first subsection that in cases where
$L < N$, recovery of the signal may still be possible if the
bandwidth of the LPF is increased.

\section{Connection with Sparse Analog Signals}
\label{sec:sparse}

In this section we depart from the subspace assumption which
prevailed until now. Instead, we would like to incorporate
sparsity into the signal model $x(t)$ of (\ref{eq:si}). To this
end, we follow the model proposed in \cite{E08} to describe
sparsity of analog signals in SI spaces. Specifically, we assume
that only $K$ out of the generators $\phi_n(t)$ are active, so
that at most $K$ of the sequences $a_n[k]$ have positive energy.

In \cite{E08}, it was shown how such signals can be sampled and
reconstructed from samples at a low rate of $2K/T$. The samples
are obtained by pre-processing the signal $x(t)$ with a set of
$2K$ sampling filters, whose outputs are uniformly sampled at a
rate of $1/T$. Without the sparsity assumption, at least $N$
sampling filters are needed where generally $N$ is much larger
than $K$. In contrast to this setup, here we are constrained to
sample at the output of a LPF with given bandwidth. Thus, we no
longer have the freedom to choose the sampling filters as we wish.
Nonetheless, by exploiting the sparsity of the signal we expect to
be able to reduce the bandwidth needed to recover $x(t)$ of the
form (\ref{eq:si}), or in turn, to reduce the number of branches
needed when using a bank of modulators.

We have seen that the ability to recover $x(t)$ depends on the
left invertibility of the matrix $\mathbf{\Phi}_{L}(\omega)$ (or
$\mathbf{\Gamma}_{L}(\omega)$). With appropriate definitions, our
problem becomes that of recovering $\hat{\ba}(e^{j\omega T})$ from
the linear set of equations (\ref{eq:meq}) (with
$\mathbf{\Gamma}_{L}(\omega)$ replacing
$\mathbf{\Phi}_{L}(\omega)$ when preprocessing is used). Our
definition of analog sparsity implies that at most $K$ of the
Fourier transforms $\hat{a}_n(\omega)$ have non-zero energy.
Therefore, the infinite set of vectors $\{\hat{\ba}(e^{j\omega
T}),\omega \in [-\pi/T,\pi/T]\}$ share a joint sparsity pattern
with at most $K$ rows that are not zero. This in turn allows us to
recover $\{\hat{\ba}(e^{j\omega T}),\omega \in [-\pi/T,\pi/T]\}$
from fewer measurements. Under appropriate conditions, it is
sufficient that $\hat{\by}(\omega)$ has length $2K$, which in
general is much smaller than $N$. Thus, fewer measurements are
needed with respect to the full model (\ref{eq:si}). The reduction
in the number of measurements corresponds to choosing a smaller
bandwidth of the LPF, or reducing the number of modulators.

In order to recover the sequences in practice, we rely on the
separation idea advocated in \cite{ME08}: we first determine the
support set, namely the active generators. This can be done by
solving a finite dimensional optimization problem under the
condition that $\mathbf{\Phi}_{L}(\omega)$ (or
$\mathbf{\Gamma}_{L}(\omega)$) are fixed in frequency up to a
possible frequency-dependent normalization sequence. Recovery is
then obtained by applying results regarding infinite measurement
vector (IMV) models to our problem \cite{ME08}. When
$\mathbf{\Phi}_{L}(\omega)$ does not satisfy this constraint, we
can still convert the problem to a finite dimensional optimization
problem  as long as the sequences $a_k[n]$ are rich \cite{E08a}.
This implies that every finite set of vectors share the same
frequency support. As our focus here is not on the sparse setting,
we do not describe here in detail how recovery is obtained. The
interested reader is referred to \cite{ME08,E08,E08a} for more
details.

The main point we want to stress here is that the ideas developed
in this paper can also be used to treat the scenario of recovering
a sparse SI signal from its lowpass content. The difference is
that now we can relax the requirement for invertibility of
$\mathbf{\Phi}_{L}(\omega), \mathbf{\Gamma}_{L}(\omega)$. Instead,
it is enough that these matrices satisfy the known conditions from
the compressed sensing literature. This in turn allows in general
reduction of the LPF bandwidth, or the number of modulators, in
comparison with the non-sparse scenario.

\section{Conclusions and Open Problems}
\label{sec:Conclusions}

This paper studied the possibility of recovering signals in SI
spaces from their low frequency components. We developed necessary
conditions on the minimal bandwidth of the LPF and sufficient
conditions on the generators of the SI space such that recovery is
possible. We also showed that proper pre-processing may facilitate
the recovery, and allow to reduce the necessary bandwidth of the
LPF. Finally, we discussed how these ideas can be used to recover
sparse SI signals from the output of a LPF.

 An
important open problem we leave to future work is the
characterization of the class of generators for which the proposed
pre-processing scheme can (or cannot) be applied. To this end, the
following question has to be answered. We formulate it only for
the most simple case of one generator (cf. also the discussion in
Example~\ref{ex:CompactSupport}).
\begin{problem}
Let $\phi \in L^{2}$ be an arbitrary function with Fourier
transform $\hat{\phi}$ and whose Grammian satisfies \eqref{eq:rc}
. Consider the shift-invariant space spanned by $\hat{\phi}$, i.e.
\begin{eqnarray*}
    \mathcal{S}_{\frac{2\pi}{T}}(\hat{\phi})
    = \cls\left\{ S_{k\frac{2\pi}{T} }\hat{\phi}\ :\ k\in\bbZ \right\}.
\end{eqnarray*}
For which functions $\phi \in L^{2}$ does there exist a function
$\hat{\gamma} \in \mathcal{S}(\hat{\phi})$ such that
$\hat{\gamma}(\omega) \neq 0$ for all $\omega\in [-\pi/T,\pi/T]$.
\end{problem}

The interesting case is when every function $\hat{\phi}(\omega -
k\, 2\pi/T)$, $k\in\bbZ$ has at least one zero in the interval
$[-\pi/T , \pi/T]$. Then the question is whether in this case it
is still possible to find a linear combination of these functions
which has no zero in $[-\pi/T , \pi/T]$.


\end{document}